\documentstyle[11pt,newpasp,twoside]{article}
\begin{document}
\title{H$\alpha$ Velocity Fields of Normal Spiral Disks}
\author{David R. Andersen$^1$, M. A. Bershady$^2$, L. S. Sparke$^2$, 
J. S. Gallagher III$^2$, E. M. Wilcots$^2$, W. van Driel$^3$,
D. Monnier-Ragaigne$^3$}
\affil{$^1$Dept. of Astronomy \& Astrophysics, Penn State University, USA\\
$^2$Department of Astronomy Department, U. Wisconsin--Madison, USA \\
$^3$Observatoire de Paris, Meudon, France}

\begin{abstract}
We present H$\alpha$ velocity fields for a sample of nearly face--on
spiral galaxies observed with DensePak on the WIYN telescope. We
combine kinematic inclinations and position angles measured from these
data with photometric inclinations and position angles measured from
$I$-band images to show that spiral disks are intrinsically non-circular.
\vskip -0.25in
\end{abstract}

\section{DensePak H$\alpha$ Velocity Fields}

DensePak is a 30$\times$45 arcsec fiber-optic integral-field unit
feeding a bench spectrograph on the WIYN 3.5m telescope (Barden et
al. 1998). By using multiple pointings of DensePak in echelle mode
($\lambda/\Delta\lambda=13,000$), we have constructed H$\alpha$
velocity fields for 8 nearby, normal, apparently face--on spiral
disks. The velocity fields, typically complete to $\sim$3 disk scale
lengths, reach the peak of the rotation curve (Figure 1), and are
modeled successfully by a single, inclined disk with a hyperbolic
tangent function for V(r) (Figure 2). Residuals from this simple model
are small (typically 5 km/s).

Our first discovery was that this sample, while selected to appear
photometrically face-on, had projected velocities indicative of
significant inclinations (up to 35$^\circ$). Clearly these disks are
not circular.

To quantify this observation, we developed an efficient method to
estimate disk elongation of nearly face-on galaxies by combining
measurements from H$\alpha$ velocity fields and $I$-band images. In
the context of a simple geometric model, we interpret differences
between kinematic and photometric inclinations and position angles in
terms of intrinsic disk ellipticity. Five galaxies in the sample are
non-circular at greater than 99\% CL; the range of ellipticities
estimated within the context of our model is 0.02 to 0.20 (Andersen et
al. 2000). Even such modest disk ellipticity can account for much of
the scatter in the Tully-Fisher (TF) relation (Franx \& de Zeeuw
1992), as discussed by Bershady \& Andersen (these proceedings).
Support for this research comes from NSF/AST-9970780.

\begin{figure}
\plotfiddle{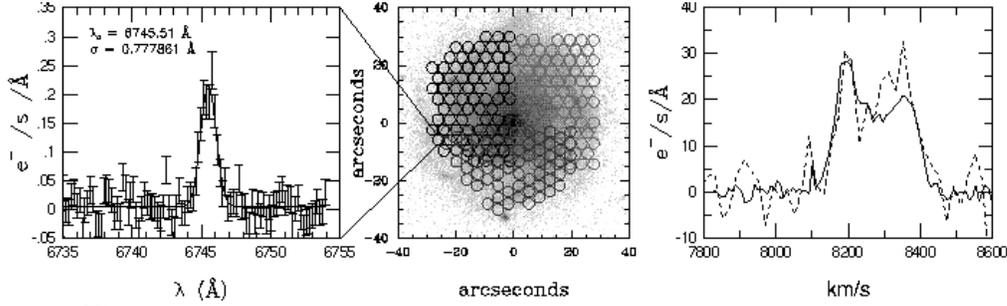}{1.25in}{-90}{54}{54}{-221}{211}
\caption{\hsize 5.25in \baselineskip 0.165in \small NGC 5123. The left panel
is a DensePak echelle spectrum at H$\alpha$ from the edge of this
nearly face-on spiral, with signal-to-noise typical of our data at 2
scale lengths. To provide adequate spatial coverage, we typically use
1 to 3 DensePak pointings to map out the velocity field (middle
panel). The right panel overlays H$\alpha$ (solid line) and HI (dashed
line) velocity profiles. The HI profile, from Nan\c{c}ay (NRT), is
normalized to the peak of the H$\alpha$ profile.  Measurements of
$W_{20}$ from these two profiles agree within errors, indicating we
have observed the peak of the optical rotation curve.}
\end{figure}
\begin{figure}
\plotfiddle{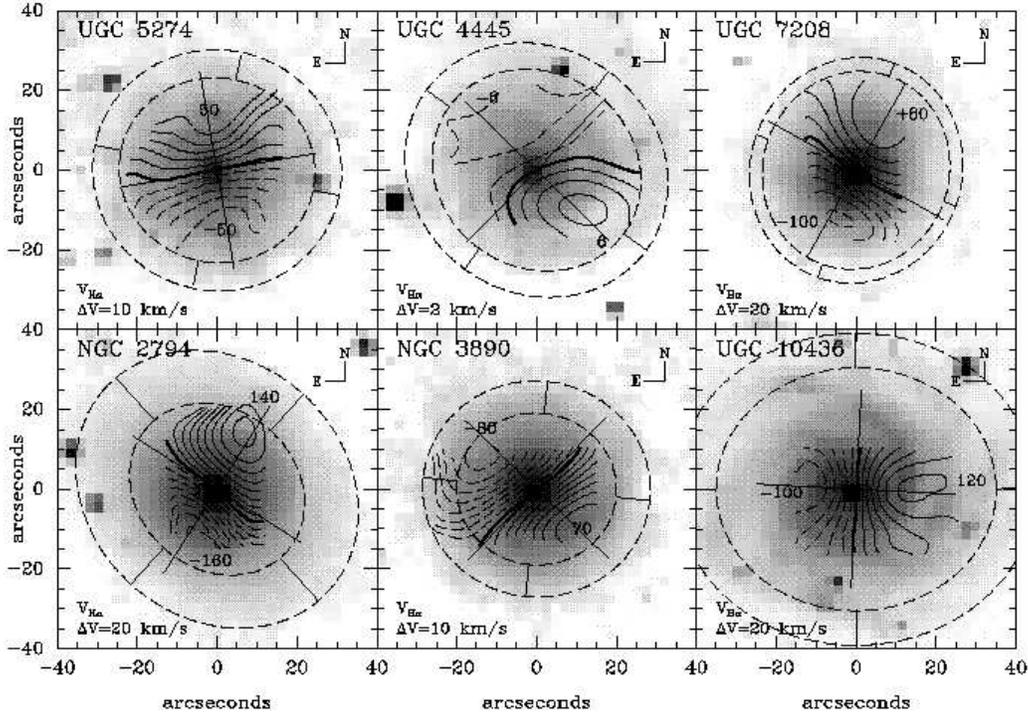}{3.3in}{0}{56}{56}{-165}{-90}
\caption{\small \hsize 5.25in \baselineskip 0.165in Polynomial surface-fits
to the DensePak H$\alpha$ velocity field for 6 sample galaxies,
overlayed on $I$-band WIYN images (NGC 5123 and UGC 4380 are presented
in Bershady \& Andersen 2000 and Andersen et al. 2000).  Solid, heavy,
and dashed lines are positive, zero, and negative velocities,
respectively, relative to the model systemic velocity. The dashed
annuli represent the radii between which photometric ${b/a}$ and PA
are measured (solid lines in annuli indicate photometric major and
minor axes). UGC 4445 is sufficiently face-on ($i<10^\circ$) that we
are unable to determine an accurate kinematic inclination, and
therefore we have excluded it from our ellipticity analysis.}
\end{figure}

\vskip -0.3in
\null

\end{document}